\documentclass [11pt, letterpaper]{article}
\usepackage{fullpage}
\usepackage{amsmath}
\usepackage{amssymb}
\usepackage{graphicx}
\usepackage{mathrsfs}
\usepackage{wrapfig}
\usepackage{boxedminipage}
\usepackage{epsfig}
\usepackage{setspace}
\usepackage{subfigure}
\usepackage{float}
\usepackage{hyperref}
\usepackage{enumerate}
\newcommand{\reef}[1]{(\ref{#1})}
\begin{document}

\begin{flushright}
\phantom{{\tt arXiv:1102.????}}
\end{flushright}

\bigskip
\bigskip
\bigskip

\begin{center} {\Large \bf Dynamics of  Fundamental Matter }
  
  \bigskip

{\Large\bf  in }

\bigskip

{\Large\bf     $\mathcal{N}=2^\ast$ Yang--Mills Theory}

\end{center}

\bigskip \bigskip \bigskip \bigskip

\centerline{\bf Tameem Albash, Clifford V. Johnson}

\bigskip
\bigskip
\bigskip

  \centerline{\it Department of Physics and Astronomy }
\centerline{\it University of
Southern California}
\centerline{\it Los Angeles, CA 90089-0484, U.S.A.}

\bigskip

\centerline{\small \tt talbash,  johnson1,  [at] usc.edu}

\bigskip
\bigskip


\begin{abstract} 
\noindent 
We study the dynamics of quenched fundamental matter in
$\mathcal{N}=2^\ast$ supersymmetric large $N$ $SU(N)$ Yang--Mills
theory at zero temperature.  Our tools for this study are probe
D7--branes in the holographically dual $\mathcal{N}=2^\ast$
Pilch--Warner gravitational background.  Previous work using D3--brane
probes of this geometry has shown that it captures the physics of a
special slice of the Coulomb branch moduli space of the gauge theory,
where the $N$ constituent D3--branes form a dense one dimensional
locus known as the enhan\c con, located deep in the infrared. Our
present work shows how this physics is supplemented by the physics of
dynamical flavours, revealed by the D7--branes embeddings we find.
The Pilch--Warner background introduces new divergences into the
D7--branes free energy, which we are able to remove with a single
counterterm. We find a family of D7--brane embeddings in the geometry
and discuss their properties.  We study the physics of the quark
condensate, constituent quark mass, and part of the meson
spectrum. Notably, there is a special zero mass embedding that ends on
the enhan\c con, which shows that while the geometry acts repulsively
on the D7--branes, it does not do so in a way that produces
spontaneous chiral symmetry breaking.

\end{abstract}
\newpage \baselineskip=18pt \setcounter{footnote}{0}

%
\section{Introduction}
%

Gauge/string dualities have emerged as powerful tools for studying the
dynamics of strongly coupled gauge theories in various regimes, some
with direct experimental interest. Some of the most striking results
have come from quite simple models such as the prototype AdS/CFT
correspondence \cite{Maldacena:1997re,Witten:1998qj,Gubser:1998bc},
where the gauge theory is ${\cal N}=4$ supersymmetric $SU(N)$
Yang--Mills theory at large $N$, with dual string (gravity) background
geometry AdS$_5\times S^5$. Despite the special simplifying properties
of this theory, studying it at finite temperature using the
gauge/gravity duality has proven (as anticipated\cite{Johnson:2003gi})
a fruitful point of contact with heavy ion collisions experiments at
RHIC in Brookhaven\footnote{Early results from the ALICE experiment at
  CERN's Large Hadron Collider seem to suggest that this overlap will
  persist to the regimes probed there as
  well\cite{Aamodt:2010pa}.}. (For a brief survey of these matters,
see refs.\cite{Johnson:2010zzb,Jacak:2010zz,Thomas:2010zzt}. For a
longer review, see for example ref.\cite{CasalderreySolana:2011us}.).

All of this should be regarded as a rough draft of the connection
between string theory studies and these kinds of experimental efforts
(albeit already an impressive and fruitful one), and if we are to fill
in more detail, much must be learned about what the essential features
are, issues of universality, and the ranges of applicability of
various computational results.  The ultimate goal would be to to
capture as much of the key features of the physics in as simple a
model as possible.  Among the important aspects to study is the fact
that the parent model is not conformally invariant once finite
temperature or density are dialed away. To capture this on the
gravity side, we must start with a model that is more complicated than
the ${\cal N}=4$ Yang Mills when at $T=0$. Such a model is the ${\cal
  N}=2^\ast$ Yang Mills model which can be obtained from the ${\cal
  N}=4$ explicitly by giving a mass to two of its adjoint scalars.

While work has been done to study the properties of the finite
temperature $\mathcal{N}=2^\ast$ theory, generalizing results obtained
for the finite temperature ${\cal N}=4$ Yang--Mills theory (see {\it
  e.g.}
refs.~\cite{Buchel:2003ah,Buchel:2004hw,Buchel:2007vy,Buchel:2008uu}),
there are no results for the dynamics of fundamental flavours in this
model. This is the focus of the present paper.  We begin the
understanding of the dynamics of fundamental quark flavours in this
model, with our eye on the issue of capturing important physical
phenomena for this sector, such as the meson spectrum and possible
thermal and chemical potential phase structure, and response to
external probes such as electromagnetic fields. This paper will focus
on the warm--up case of zero temperature, which is already very
interesting due to the complexity of the dual geometry.

The $\mathcal{N}=2^\ast$ supersymmetric Yang--Mills theory is realized by giving equal masses
to two chiral multiplets.  In terms of the $\mathcal{N} = 4$ theory,
the deformation involves two operators, a bosonic one of dimension
two, and a fermionic one of dimension three.  This is captured
holographically by the Pilch--Warner (PW) flow of
ref.~\cite{Pilch:2000ue}, where the massive deformation is captured in
the holographic dual by two scalar fields $\alpha$ (dual to the
$\Delta = 2$ operator) and $\chi$ (dual to the $\Delta = 3$ operator)
in five dimensional, gauged $\mathcal{N}=8$ supergravity. It was shown in
refs.~\cite{Buchel:2000cn,Evans:2000ct}, using a D3--brane analysis,
that this flow geometry captures a very specific point of the Coulomb
branch moduli space. It is a particularly clear example of how the
enhan\c con mechanism~\cite{Johnson:1999qt} resolves physics that
appears singular in supergravity. The complete physics is established
using the physics of the probes, and is consistent with known gauge
theory phenomena: the enhan\c con itself is a large $N$ manifestation
of the Seiberg--Witten \cite{Seiberg:1994rs,Seiberg:1994aj} locus at the origin of the Coulomb branch,
where the constituent branes of the geometry have spread into a dense
one dimensional locus of points of length set by the ${\cal N}=4$
breaking mass.

Adding fundamental matter corresponds to introducing D7--branes to the
background~\cite{Karch:2002sh}, which is dual to introducing a $\Delta
= 3$ operator on the field theory side.  Our D7--branes will only be
probes, meaning that the D7--branes do not back--react on the gravity
background formed by the D3--branes, which corresponds to the
fundamental matter being in the ``quenched'' limit.  This is
accomplished by having the number of D7--branes $N_f$ be much less
than the number $(N)$ of D3--branes.  In fact, we will take $N_f = 1$.

Even though we restrict ourselves in this work to zero temperature,
the D7--brane embeddings are non--trivial and reveal some interesting
physics.  Since the background contains a dilaton, axion, and NS--NS
and R--R forms (we review the gravity background in sections
\ref{sec:gravityA} and~\ref{sec:gravityB}), the D7--brane action is
quite involved.  In section \ref{sec:D7}, we find a consistent ansatz
for the transverse fields to the D7--brane that allows us to restrict
our analysis to the Dirac--Born--Infeld (DBI) action for the D7--brane
with only the dilaton turned on.  Even with this much simplified
scenario, we show a new ultraviolet (UV) divergence in the action of
the D7--brane, and we propose a counterterm to eliminate it.  We show
analytically that with this counterterm, the condensate as a function
of bare quark mass behaves as expected for large mass (\emph{i.e.} it
vanishes), suggesting that this is indeed the correct counterterm.  We
then numerically study the condensate (vev of the operator) as a
function of bare quark and show that it scales as it should in terms
of the mass of the massive deformation of the underlying
theory. Section~\ref{sec:ConMass} presents our study of the
constituent quark mass, and in section \ref{sec:meson} we study the
fluctuation of one of the transverse directions of the D7--brane (this
is dual to calculating the meson spectrum \cite{Karch:2002xe}),
showing that the mass of the dual meson is above that of mesons of
$\mathcal{N}=4$ supersymmetric Yang--Mills.  Our meson spectrum
analysis is not complete as we do not calculate the fluctuations along
the other transverse direction, but the form of the $C_{(6)}$ and
$C_{(8)}$ is not known and hence makes this analysis not possible at
this time. We conclude in section~\ref{sec:conclusions}.

%
\section{Review} \label{sec:review}
%
\subsection{Gauge Theory}
%
The matter content of the $\mathcal{N}=4$ supersymmetric Yang--Mills
theory consists of a gauge multiplet containing the bosonic fields
$(A_\mu, X_j)$, $j=1,\dots, 6$, where $X_j$ are real scalars
transforming as the $\mathbf{6}$ of $SO(6)$, and the fermionic fields
$(\lambda_j)$, $j = 1, \dots, 4$, which transform as the $\mathbf{4}$ of
$SU(4)$.  Writing this matter content in terms of $\mathcal{N}=1$
superfields, the theory has one vector supermultiplet $(A_\mu,
\lambda_4)$ and three chiral multiplets $\Phi_j = (\lambda_j, \phi_j =
X_{2j - 1} + i X_{2 j})$, $j = 1, \dots, 3$.

We can make two of the
chiral multiplets massive (with equal mass), preserving only an ${\cal N}=2$:
\begin{equation}
\mathcal{L}_{\mathcal{N}=2} = \mathcal{L}_{\mathcal{N}=4} + \int d^2 \theta  \left( \frac{1}{2} m \Phi_1^2 + \frac{1}{2} m \Phi_2^2 \right) \ .
\end{equation}
This choice preserves an $SU(2) \times U(1)$ subgroup of the
original $SO(6)$ $R$--symmetry.  For us, we focus on a very specific
combination of the fields that will be massive:
\begin{eqnarray}
\alpha: & & \sum_{i=1}^4 \mathrm{Tr}(X_i X_i) - 2 \sum_{i=5}^6 \mathrm {Tr}(X_i X_i) \ ,\\
\chi: && \mathrm {Tr}(\lambda_1 \lambda_1 + \lambda_2 \lambda_2 ) + \mathrm{h.c.} \ .
\end{eqnarray}
In the dual gravity picture, this will correspond to turning on two
real scalars $\alpha$ and $\chi$ with conformal dimension
$\Delta_\alpha =2$ and $\Delta_{\chi}=3$ respectively. The theory
described by the resulting flow to the infrared (IR) is called the
${\cal N}=2^\ast$ theory.

Giving a vacuum expectation value for $\phi_3$ sets the potential
$[\phi_3, \phi_3^\dagger]^2$ to zero, and corresponds to moving on the
accessible part of the Coulomb branch moduli space of vacua, where the
$SU(N)$ gauge theory is broken to $U(1)^{N-1}$. The moduli space is an
$N-1$ complex dimensional space, in general, although in the gravity
dual, only a single complex dimensional subspace is explicit, as we will
review below\cite{Buchel:2000cn,Evans:2000ct}.
%

\subsection{Five--Dimensional Supergravity Solution} \label{sec:gravityA}
%
In this section, we briefly review the construction of the
Pilch--Warner background, starting with a solution to $\mathcal{N}=8$
five dimensional gauged supergravity. The lift to ten
dimensions will be discussed in the next section.

The relevant part of  the five dimensional gauged supergravity
action is given by:
\begin{equation}
S_{5d} = \frac{1}{16 \pi G_{5}} \int d^5 x \sqrt{-G} \left( \mathcal{R} - 12 \left( \partial \alpha \right)^2 - 4 \left( \partial \chi \right)^2 - 4 \mathcal{P} \right) \ ,
\end{equation}
where the potential $\mathcal{P}$ for the scalars $\chi$ and $\alpha$ is given by:
\begin{equation}
\mathcal{P} = \frac{g^2}{16} \left[ \frac{1}{3} \left( \frac{ \partial W}{\partial \alpha} \right)^2 + \left( \frac{\partial W}{\partial \chi} \right)^2 \right] - \frac{g^2}{3} W^2 \ .
\end{equation}
The constant $g$ is related to the AdS radius $R$, $g^2 = 4 / R^2$, and the superpotential $W$ is given by:
\begin{equation}
W = - e^{- 2 \alpha} - \frac{1}{2} e^{4 \alpha} \cosh\left( 2 \chi \right) \ .
\end{equation}
The equations of motion derived from this action are given by:
\begin{equation*}
\mathcal{R}_{\mu \nu} - 12 \partial_\mu \alpha \partial_\nu \alpha - 4 \partial_\mu \chi \partial_\nu \chi - \frac{4}{3} G_{\mu \nu} \mathcal{P} = 0 \ ,
\end{equation*}
\begin{equation}
\frac{1}{\sqrt{-G}} \partial_\mu \left( \sqrt{-G} G^{\mu \nu} \partial_\nu \alpha \right) - \frac{1}{6} \frac{\partial \mathcal{P}}{\partial \alpha} = 0 \ , \quad \frac{1}{\sqrt{-G}} \partial_\mu \left( \sqrt{-G} G^{\mu \nu} \partial_\nu \chi \right) - \frac{1}{2} \frac{\partial \mathcal{P}}{\partial \chi} = 0 \ ,
\end{equation}
where we have used that the Ricci scalar is given by:
\begin{equation}
\mathcal{R} = 4 \left( \frac{5}{3} \mathcal{P} + 3 \left( \partial \alpha \right)^2 + \left( \partial \chi \right)^2 \right)  \ .
\end{equation}
We consider a metric ansatz of the form:
\begin{equation}
ds_5^2 = e^{2 A} \left( - dt^2 + d \vec{x}^2 \right) + dr^2 \ , 
\end{equation}
and for a supersymmetric solution, the problem of solving for
$\chi(r)$, $\alpha(r)$ and $A(r)$ reduces to solving the following
first order equations \cite{Pilch:2000ue}:
\begin{eqnarray}
  \frac{d \rho}{d r} &=& \frac{\rho}{3 R} \left( \frac{1}{\rho^2} - \rho^4 \cosh( 2 \chi ) \right) \ , \nonumber \\
  \frac{d A}{d r} &=& \frac{2}{3 R} \left( \frac{1}{\rho^2} + \frac{1}{2} \rho^4 \cosh( 2 \chi ) \right) \ , \nonumber \\
  \frac{d \chi}{d r} &=& -\frac{1}{2 R}  \rho^4 \sinh( 2 \chi ) \ , \label{eq:firstorder}
\end{eqnarray}
where $\alpha(r) = \ln \rho(r)$.  Partial solutions are given by \cite{Pilch:2000ue}:
\begin{equation}
e^A = \frac{ k \rho^2}{\sinh(2 \chi)} \ , \quad \rho^6 = \cosh(2 \chi) + \sinh^2 (2 \chi) \left( \ln \left( \tanh(\chi) \right) + \gamma \right) \ ,
\end{equation}
where $k = m R$, the mass of the chiral multiplets dual to the fields
$\alpha$ and $\chi$, and $\gamma$ is a constant, which we choose to be
less than or equal to zero for the duration of our work.  Different
values for $\gamma \leq 0$ correspond to different slices through the
moduli space of $\mathcal{N}=2^\ast$ in the IR, with $\gamma = 0$
describing a singular point \cite{Pilch:2000ue,Gubser:2000nd}, the
physics of which was uncovered in
refs.\cite{Buchel:2000cn,Evans:2000ct}, as we will review below. We
can solve for $\chi$ numerically, and it is convenient to do so in a
coordinate system given by:
\begin{equation}
\frac{z}{R} = \hat{z} = e^{-r / R} \ . \label{eq:newz}
\end{equation}
We will restrict the ``hat'' coordinates to be dimensionless
coordinates.  Near the AdS boundary, the fields have an expansion given by:
\begin{eqnarray}
\chi &=& k \hat{z} \left[ 1 + k^2 \hat{z}^2 \left( \frac{1}{3}(1+4\gamma) + \frac{4}{3} \ln (k \hat{z}) \right) + k^4 \hat{z}^4 \left( \frac{1}{90}(-7+300\gamma+200\gamma^2) + \frac{10}{9}(3 + 4\gamma) \ln (k \hat{z}) \right. \right. \nonumber \\
&& \left. \left. + \frac{20}{9} \ln(k \hat{z})^2 \right) + \mathcal{O} (k^6 \hat{z}^6 \ln(k \hat{z})^3) \right] \ , \nonumber \\
\rho &=& 1 + k^2 \hat{z}^2 \left( \frac{1}{3}(1+2 \gamma) + \frac{2}{3} \ln( k \hat{z}) \right) + k^4 \hat{z}^4 \left( \frac{1}{18}(1+3 \gamma+12\gamma^2) + (2+ \frac{4}{3}\gamma) \ln (k \hat{z}) \right. \nonumber \\
&& \left. + \frac{2}{3} \ln(k \hat{z})^2 \right) + \mathcal{O} ( k^6 \hat{z}^6 \ln( k \hat{z})^3 ) \ , \nonumber \\
A &=& - \ln (2 \hat{z}) - \frac{1}{3} k^2 \hat{z}^2 - k^4 \hat{z}^4 \left( \frac{2}{9} \left( 1+ 5 \gamma+ 2 \gamma^2 \right) + \frac{2}{9} \left(5 + 4 \gamma \right) \ln(k \hat{z}) + \frac{4}{9} \ln( k \hat{z})^2 \right) \\ \nonumber 
&&+ \mathcal{O} (k^6 \hat{z}^6 \ln( k \hat{z})^3) \ .
\end{eqnarray}

For $\gamma = 0$, in the IR at $\hat{z} = \infty$,  $\chi$ diverges and $\rho$
vanishes.  For $\gamma<0$, $\chi$ remains finite
and $\rho$ vanishes at $\hat{z} = \infty$.  To analyze the behavior
near $\hat{z} = \infty$, it is convenient to make the following 
coordinate change:
\begin{equation}
r' = R \sqrt{\frac{c+1}{c-1}} \ ,
\end{equation}
where $c = \cosh(2 \chi)$.  The coordinate $r'$ ranges from $r_0' $
(the IR) to infinity (the UV), where for $\gamma = 0$, $r_0' = R$
while for $\gamma < 0$, $r_0' > R$.  We will sometimes refer to the
radius $r_0'$ as the ``singular locus''.
%
\subsection{Ten Dimensional Supergravity Uplift}  \label{sec:gravityB}
%
Solutions of the five dimensional truncation of the previous section
may be uplifted or oxidized to give  a solution of type IIB supergravity
\cite{Pilch:2000ue}, with the ten dimensional metric (in Einstein
frame) given by:
\begin{equation}
ds_{10}^2 = \Omega^2 d s_5^2+ \frac{a^2}{2} \frac{ \Omega^2}{\rho^2} \left(c^{-1} d\theta^2 + \rho^6 \cos^2 \theta \left(\frac{\sigma_1^2}{c X_2} + \frac{\sigma_2^2 + \sigma_3^2}{X_1} \right) + \sin^2 \theta \frac{d \phi^2}{X_2} \right) \ ,
\end{equation}
where:
\begin{equation} \begin{array}{rcl}
X_1(r, \theta) &=& \cos^2 \theta + \rho(r)^6 \cosh(2 \chi(r)) \sin^2 \theta \ , \\
X_2(r, \theta) &=& \cosh(2 \chi(r)) \cos^2 \theta + \rho(r)^6 \sin^2 \theta \ , \\
\Omega^2 &=& \frac{\left(c X_1 X_2\right)^{1/4}}{\rho} \ , \quad
a^2 = \frac{8}{g^2} = 2 R^2 \ , \\
c &=& \cosh(2 \chi) \ , \quad
\rho =  e^{\alpha} \ , 
\end{array}
\end{equation}
and there is a deformed $S^3$ with $SU(2)$ invariant 1--forms
\begin{equation}
\begin{array}{rcl}
  \sigma_1 &=& \frac{1}{2} \left( d\alpha + \cos \psi d \beta \right) \ , \\
  \sigma_2 &=& \frac{1}{2} \left( - \sin \alpha d \psi + \cos \alpha \sin \psi d \beta \right) \ , \\
  {\rm and}\quad \sigma_3 &=& \frac{1}{2} \left( \cos \alpha d \psi + \sin \alpha \sin \psi d \beta \right) \ .
\end{array}
\end{equation}
Using the conventions of ref.~\cite{Buchel:2000cn} for the ten dimensional fields, the dilaton and axion field are encoded in a complex function $B$ given by:
\begin{equation}
C_{(0)} + i e^{-\Phi} = i \frac{1 + B}{1-B} \ .
\end{equation}
The NS--NS two form and the RR two--form are encoded in a single complex two--form $A_{(2)}$ given by:
\begin{equation}
A_{(2)} = C_{(2)} + i B_{(2)} \ .
\end{equation}
The solutions in terms of these complex fields is given by:
\begin{equation}
B = e^{2 i \phi }\frac{ \sqrt{c X_1} - \sqrt{X_2}}{\sqrt{c X_1} + \sqrt{X_2}} \ ,
\end{equation}
and\footnote{We are using a slightly different notation than in ref.~\cite{Pilch:2000ue} by pulling out the $i$'s and their sign.}:
\begin{equation}
A_{(2)} = e^{i \phi} \left( - i a_1 d \theta \wedge \sigma_1 + i  a_2 \sigma_2 \wedge \sigma_3 + a_3 \sigma_1 \wedge d \phi \right) \ ,
\end{equation}
with
\begin{eqnarray}
a_1 &=& \frac{4}{g^2} \tanh(2 \chi) \cos \theta \ , \nonumber \\
a_2 &=& \frac{4}{g^2} \frac{\rho^6 \sinh(2 \chi)}{X_1} \sin \theta \cos^2 \theta \ , \nonumber \\
{\rm and}\quad  a_3 &=& \frac{4}{g^2} \frac{\sinh(2 \chi)}{X_2} \sin \theta \cos^2 \theta \ .
\end{eqnarray}
Therefore, extracting the real fields of ten dimensional supergravity, we have:
\begin{equation}
C_{(0)} = - \frac{\sin\phi \cos\phi (c X_1 - X_2) }{ c X_1 \sin^2 \phi + X_2 \cos^2 \phi} \ , \quad
e^{-\Phi}  = \frac{\sqrt{c X_1 X_2}}{ c X_1 \sin^2 \phi + X_2 \cos^2 \phi} \ ,
\end{equation}
and
\begin{eqnarray}
B_{(2)} &=& - \cos \phi a_1 d \theta \wedge \sigma_1 + \cos \phi a_2 \sigma_2 \wedge \sigma_3 +  \sin \phi a_3 \sigma_1 \wedge d\phi  \ , \\
C_{(2)} &=& \sin \phi a_1 d \theta \wedge \sigma_1 -  \sin \phi  a_2 \sigma_2 \wedge \sigma_3 + \cos \phi a_3  \sigma_1 \wedge d\phi \ .
\end{eqnarray}
The antisymmetric 5--form tensor field strength is given by:
\begin{equation}
F_{(5)} = \mathcal{F} + \ast \mathcal{F} \ ,
\end{equation}
where\footnote{We are using the factor of 4 convention from ref.~\cite{Buchel:2000cn,Evans:2000ct} and not the conventions of ref.~\cite{Pilch:2000ue}.}
\begin{equation}
\mathcal{F} =4 dx^0 \wedge dx^1 \wedge dx^2 \wedge dx^3 \wedge d w  \ .
\end{equation}
and 
\begin{equation}
w =\frac{k^4}{4 g_s}  \left( \frac{\rho^6 X_1}{\sinh^4 (2 \chi)}\right) \ .
\end{equation}
Using that $F_{(5)} = d C_{(4)}+C_{(2)} \wedge d B_{(2)}$, we can write the four--form potential as:
\begin{equation}
C_{(4)} = 4 w dx^0 \wedge dx^1 \wedge dx^2 \wedge dx^3 + \alpha_4 \ , 
\end{equation}
where $\alpha_4$ is a 4--form with legs in the 6 transverse dimensions, satisfying:
\begin{equation}
\ast \left(4 d w \wedge d x^0 \wedge d x^1 \wedge d x^2 \wedge d x^3 \right) =  d \alpha_4 + C_{(2)} \wedge d B_{(2)} \ .
\end{equation}
Since we have non--zero $C_{(0)}$ and $C_{(2)}$, there is a non--zero $C_{(6)}$ and $C_{(8)}$.  These can be calculated \emph{via}:
\begin{equation}
\ast d C_{(0)} = d C_{(8)} + C_{(6)} \wedge d B_{(2)} \ , \quad \ast \left( d C_{(2)} + C_{(0)} d B_{(2)} \right) = - \left( d C_{(6)} + C_{(4)} \wedge d B_{(2)} \right) \ .
\end{equation}
We do not have analytic forms for these expressions, but it will be important for us to calculate the $\phi$ dependence of these two terms.  First, we write:
\begin{equation}
B_{(2)} = \cos \phi \ b_2 + \sin \phi \ d \phi \wedge  b_1 \ , \quad C_{(2)} = \sin \phi \ c_2 + \cos \phi \ d \phi \wedge  c_1 \ , 
\end{equation}
\begin{equation}
d B_{(2)} = \sin \phi \  d \phi \wedge h_2 + \cos \phi \ h_3 \ , 
\end{equation}
where $(b_1,c_1)$, $(b_2, c_2,h_2)$ and $h_3$ are one--forms,
two--forms, and a three form respectively with no legs along $\phi$.
Also:
\begin{equation}
\ast d C_{(2)} = \cos \phi \ g_7 + \sin \phi \ d\phi \wedge g_6 \ ,
\end{equation}
where $g_7 $ is a seven--form with no legs along $\phi$ and $g_6$ is a
six--form with a leg along $\phi$.  Also, we have:
\begin{equation}
\ast d C_{(0)} =  \frac{ X_2 \cos^2\phi - c X_1 \sin^2 \phi}{\left(c X_1 \sin^2\phi + X_2 \cos^2 \phi \right)^2} g_9 + \left( \frac{\sin \phi \cos \phi}{c X_1 \sin^2 \phi + X_2 \cos^2 \phi} \right) d \phi \wedge g_8 \ , 
\end{equation}
\begin{equation}
\ast C_{(0)} d B_{(2)} = \frac{\sin \phi \cos \phi}{c X_1 \sin^2 \phi + X_2 \cos^2 \phi} \left( \sin \phi \ h_7 + \cos \phi \ d \phi \wedge h_6 \right) \ .
\end{equation}
Using that $\alpha_4 \wedge d B_{(2)} = 0$, we can write:
\begin{equation} \label{eqt:dC6}
d C_{(6)} = \sin \phi  \ d \phi \wedge f_6 + \cos \phi \ f_7 - \frac{\sin \phi \cos \phi}{c X_1 \sin^2 \phi + X_2 \cos^2 \phi} \left( \sin \phi \ h_7 + \cos \phi \ d \phi \wedge h_6 \right) \ .
\end{equation}
In principle, we can write a general form for $C_{(6)}$ as follows \cite{Buchel:2000cn}:
\begin{equation} \label{eqt:C6} C_{(6)} = \left[ f_2(x) + f_1(x)
    \wedge d \phi \right] \wedge dx^0 \wedge dx^1 \wedge dx^2 \wedge
  dx^3 \ ,
\end{equation}
where $f_2$ and $f_1$ do not have legs along $\phi$ but may depend on
$\phi$.  From the form of equation~\reef{eqt:dC6}, $f_2(x)$ must
depend on $\phi$.  Furthermore, it is not difficult to see from
equation~\reef{eqt:dC6} that $f_2(x)$ will be proportional to $\cos
\phi$ or a function $F(\cos \phi)$ that vanishes at $\cos \phi = 0$.

Finally, let us consider $C_{(8)}$.  Its equation can be written in the form:
\begin{eqnarray} \label{eqt:dC8}
d C_{(8)} &=&  \frac{ X_2 \cos^2\phi - c X_1 \sin^2 \phi}{\left(c X_1 \sin^2\phi + X_2 \cos^2 \phi \right)^2} g_9 + \left( \frac{\sin \phi \cos \phi}{c X_1 \sin^2 \phi + X_2 \cos^2 \phi} \right) d \phi \wedge g_8 \\
&& + \left( \sin \phi f_2 \wedge d \phi \wedge h_2 + \cos\phi f_2 \wedge h_3 + \cos \phi f_1 \wedge d \phi \wedge h_3 \right) \wedge (dx)^4 \ .
\end{eqnarray}
Again, we can write the most general form for $C_{(8)}$ as:
\begin{equation} \label{eqt:C8}
C_{(8)} = \left[ f_4(x) + f_3(x) \wedge d \phi \right]  \wedge dx^0 \wedge dx^1 \wedge dx^2 \wedge
  dx^3  \ ,
\end{equation}
where $f_4$ and $f_3$ do not have legs along $\phi$ but may depend on
$\phi$.  As before, $f_4(x)$ must depend on $\phi$ based on equation
\reef{eqt:dC8}.  In particular, what will be of use to us later is the
remark that $\delta_\phi f_4 \propto \cos \phi$ and/or a function $G(\cos \phi)$ that vanishes at $\cos \phi = 0$.
Furthermore, $ f_3 $ is  proportional to $\cos \phi$ or a function 
$H(\cos\phi)$ that vanishes at $\cos \phi = 0$.
%

\subsection{Probe D3--branes and the Coulomb Branch}
%
Let us briefly review some of the physics uncovered in
refs.~\cite{Buchel:2000cn,Evans:2000ct}.  The authors probed the ten dimensional supergravity background with a D3--brane extended along
the $(t,x_1,x_2,x_3)$ directions.  This probe can explore only a single
complex dimensional subspace of the gauge theory's moduli space (of
the Coulomb branch), and this corresponding to the plane $(r',\phi)$
at $\theta = \pi/2$.  Moving the probe out of this plane results in a
non--zero potential.  Despite the supergravity background being
singular at $r' = r'_0$, the metric on the D3--brane probe is regular.
For $\gamma =0$, the tension of the D3--brane vanishes at $r'_0=
r'_e$.  This locus of points on the Coulomb branch is called the
enhan\c con\cite{Johnson:1999qt}, and this is where new physics, not
visible in supergravity analysis alone, appears. The $N$ constituent
D3--branes of the supergravity solution are tensionless there and
delocalize, spreading out into a dense locus of points: the
$\theta=\pi/2$ circle at $r'=r'_e$ in the $(r',\phi)$ plane. The
spacetime ends at $r'=r'_e$. In fact, in variables chosen so as to
normalize the probe theory's kinetic terms
appropriately\cite{Buchel:2000cn}, the circle becomes a cut running
from $-k$ to $k$, where $k=mR$. This is a large $N$ Seiberg--Witten
locus, with its size set by the natural scale $m$, the mass of the
adjoint scalars that break ${\cal N}=4$ to ${\cal N}=2$. For
$\gamma<0$, the tension remains finite at $r'_0 (> r'_e)$ and is
associated with the dyons having a finite mass.  For $\gamma > 0$, the
tension is negative at $r'_0 (< r'_e)$ suggesting that the supergravity
background with $\gamma >0$ are unphysical.

\section{Fundamental Flavours and D7--Branes} \label{sec:D7}
%
\subsection{A Consistent Embedding}
Let us now introduce fundamental matter into the $\mathcal{N}=2^\ast$
gauge theory.  This is done by introducing D7--branes into the dual
gravity background \cite{Karch:2002sh}.  In the field theory, this
amounts to introducing an operator $\mathcal{O}_q$ with source $m_q$
given by \cite{Kobayashi:2006sb}:
\begin{equation}
\mathcal{O}_q = i \tilde{\psi} \psi + \tilde{q} \left(m_q + \sqrt{2} \phi_3 \right) \tilde{q}^{\dagger} + q^{\dagger} \left( m_q + \sqrt{2} \phi_3 \right) q + \mathrm{h.c.} \ .
\end{equation}
The D7--brane action is given by:
\begin{equation}
S_{D7} = - \mu_7 \int_{\mathcal{M}_8} d^8 \xi \ e^{-\Phi} \sqrt{ - \det \left( P \left[ G^{(s)} + B \right]_{ab} + 2 \pi \alpha' F_{ab} \right)} + \mu_7 \int_{\mathcal{M}_8} P\left[e^{2 \pi \alpha' F + B} \wedge \oplus_n C_n \right]  \ ,
\end{equation}
where $G^{(s)}_{\mu \nu}$ is the string frame metric.  Note that for
the general Pilch--Warner solution, the Wess--Zumino (WZ) sector will
include terms containing the $C_{(4)}$ potential, the $C_6$ potential,
the 2--form $B$ field, and the $C_{(8)}$ potential:
\begin{equation}
\int  \left(P \left[B_{(2)} \right] + 2\pi \alpha' F \right)^2 \wedge C_{(4)} \ , \quad \int \left(P \left[B_{(2)} \right] + 2\pi \alpha' F \right) \wedge P \left[C_{(6)} \right] \ , \quad \int P\left[ C_{(8)} \right] \ .
\end{equation}
We wish to consider a D7--brane embedding in static gauge, with four
coordinates in the D3--brane ``gauge theory directions'' ($x^0,\ldots,
x^3$), three wrapped on the (deformed) $S^3$, and transverse to $\phi$
and $\theta$:
\begin{equation}
\xi^a = x^a \ , a = 0, \dots, 7 \ , \quad \phi \equiv \phi(z) \  , \quad \theta \equiv \theta(z) \ .
\end{equation}
We will argue that there is a consistent choice for $\phi(z)$ where it
is a constant.  Note that contributions from the WZ terms to the
equation of motion for $\phi$ will come from terms like:
\begin{equation}\begin{array}{c}
P\left[B_{(2)} \right] \wedge { \delta_\phi P\left[ B_{(2)} \right]}{ } \wedge  P \left[C_{(4)} \right] \ , \quad P\left[B_{(2)} \right]^2 \wedge { \delta_\phi P \left[C_{(4)} \right] }{ }  \ ,  \\ \\
 \quad P\left[B_{(2)} \right] \wedge { \delta_\phi P \left[C_{(6)} \right] }{ } \ , \quad { \delta_\phi P\left[ B_{(2)} \right]}{ } \wedge P \left[C_{(6)} \right]   \ , \quad { \delta_\phi P \left[C_{(8)} \right] }{ } \quad etc.  
\end{array}
\end{equation}
We note that $B_{(2)} = 0$ for $\phi = (2 n + 1) \pi /2$.  Therefore,
all the terms involving $C_{(4)}$ vanish.  The first perhaps
non--trivial term to check is the one involving ${ \delta_\phi P\left[
    B_{(2)} \right]}{ } \wedge P \left[C_{(6)} \right]$.  From our
arguments resulting in equation \reef{eqt:C6}, we can argue that:
\begin{equation}
P \left[ C_{(6)} \right] \Big|_{\phi = (2 n + 1) \pi /2}  = 0 \ .
\end{equation}

We now need to check whether the last term is zero as well. From our
arguments around equation \reef{eqt:C8}, we have that:
\begin{equation}
\delta_\phi P \left[ C_{(8)} \right] \Big|_{\phi = (2 n + 1) \pi /2}  = 0 \ .
\end{equation}

Finally, the variation of the dilaton with respect to $\phi$ will also
be proportional to $\sin\phi \cos\phi$ and so its contribution to the
equation of motion will be zero as well.  Therefore, these results
suggest that we can consistently take $\phi = (2n+1)\pi / 2$.  With
this choice, we can also consistently take $F = 0$ since all the
source terms in the equation of motion for $F$ will vanish at this
value of~$\phi$ (this includes the terms linear in $\alpha'$ that
appear from expanding the DBI action and from the $F_{(2)} \wedge
P[C_{(6)}]$ term).  Finally, since $C_{(0)}$ vanishes at this value of
$\phi$, so will $C_{(8)}$, and we can write an effective action for
the D7--brane embedding with only the transverse field $\theta(z)$ as:
\begin{equation}
S_{D7} = - \mu_7 \int_{\mathcal{M}_8} d^8 \xi \ e^{\Phi} \sqrt{ - \det \left( P \left[ G \right]_{ab} \right)}  \ ,
\end{equation}
where we have converted our metric to Einstein frame and the dilaton is simply given by:
\begin{equation}
e^{\Phi} = \sqrt{\frac{c X_1}{X_2}} \ .
\end{equation}
Substituting  everything in, we can write the action as:
\begin{equation}
S_{D7} =  - 2 \pi^2 R^3 \mu_7 \int d^4 x \int d z \ c(z) \rho(z)^2   X_1(\rho(z), \chi(z), \theta(z))^{1/2} \cos^3 \theta e^{4 A(z)} \sqrt{  \frac{R^2}{z^2} + \frac{R^2}{c(z) \rho(z)^2} \theta'(z)^2} \ . \label{eq:actionD7}
\end{equation}

\subsection{Regularizing the Action} \label{sec:counterterm} As is by
now standard procedure in the literature (see {\it{e.g.}} refs.~\cite{Karch:2005ms,Karch:2006bv}), the D7--brane action,
UV divergent as one approaches the AdS boundary, must be regularized
before physics can be extracted from it. There are standard
counterterms that must be added, that appear here, and in addition we
find a new one that arises in the case in hand.

It is convenient again to use the dimensionless coordinate $\hat{z}$,
defined in equation~\reef{eq:newz}.  From our
action~\reef{eq:actionD7}, we can calculate the equation of motion for
$\theta(\hat{z})$, which turns out to have asymptotic behaviour given
by:
\begin{equation} \label{eqt:asymptotics}
\theta(\hat{z}) = \hat{z} \left( \theta_0 + \hat{z}^2 \theta_2 - \frac{2}{3} k^2 \theta_0 \hat{z}^2 \ln(\hat{z}) + \dots \right) \ .
\end{equation}
Notice that the appearance
of the logarithmic behavior is striking. It is worth noting that it is
\emph {not} due to a non--zero Ricci scalar in the four dimensional
space as in refs.~\cite{Karch:2005ms,Karch:2006bv}. We will find a new
source for this behaviour in this background.

Note that  the factor $R
\exp(4 A(\hat{z}))/ \hat{z}$ in equation \reef{eq:actionD7} is simply the square root of the
determinant of the metric in the five dimensional picture.  It has
expansion:
\begin{equation}
\frac{R}{\hat{z}} e^{4 A(\hat{z})} = \frac{R}{\hat{z}} \left( \frac{1}{16 \hat{z}^4} - \frac{k^2}{12 \hat{z}^2} - \frac{k^4}{18} \left( 5 \ln(k \hat{z}) + 2 \ln (k \hat{z})^2 \right) + \mathcal{O}(\hat{z}^2) \right) \ .
\end{equation}
We can interpret the $c(\hat{z}) \rho(\hat{z})^2$ and the $X_1(\rho,
\chi, \theta)^{1/2}$ in equation \reef{eq:actionD7} as new interaction terms.  To understand the
divergences of this action, we expand the action
to $\mathcal{O}(\hat{z}^4)$:
\begin{eqnarray}
S_{D7} &=&  - 2 \pi^2 R^3 \mu_7 \int d^4 x \int d \hat{z} \ \frac{R}{\hat{z}} e^{4 A(\hat{z})} \left( 1 + 2 \alpha(\hat{z}) + 2 \chi(\hat{z})^2 + \frac{2}{3} \chi(\hat{z})^4+ 2 \alpha(\hat{z})^2 + 4 \alpha(\hat{z}) \chi(\hat{z})^2 \right. \nonumber \\
&&\left.+ \frac{\hat{z}^2}{2} \theta'(\hat{z})^2 \left( 1- \frac{3}{2} \theta(\hat{z})^2 \right) - \frac{\hat{z}^4}{8} \theta'(\hat{z})^4 - \frac{3}{2} \theta(\hat{z})^2 + \frac{7}{8} \theta(\hat{z})^4 -2 \chi(\hat{z})^2 \theta(\hat{z})^2 \right) \ .
\end{eqnarray}
Integrating and keeping only divergent terms, we have:
\begin{equation}
S_{D7} =  -2 \pi^2 R^4 \mu_7 \int d^4 x \left(  \frac{1}{64 \hat{\epsilon}^4} + \frac{k^2}{16 \hat{\epsilon}^2} + \frac{k^2 \ln(k\hat{\epsilon})}{24 \hat{\epsilon}^2} - \frac{7 k^4 \ln(k\hat{\epsilon})^2}{36}  - \frac{\theta_0^2}{32 \hat{\epsilon}^2} + \frac{k^2 \theta_0^2 \ln(\hat{\epsilon})}{12} +\mathcal{O}(\hat{\epsilon}^0) \right) \ ,
\end{equation}
In particular, the new divergence associated with $\theta_0$ comes
from the following terms:
\begin{equation}
\int d\hat{z} \ \left( \left(\frac{1}{16 \hat{z}^5} - \frac{k^2}{12 \hat{z}^3} \right) \frac{\hat{z}^2}{2} \theta'(\hat{z})^2 + \frac{1}{16 \hat{z}^5} \left(- 2 \chi(\hat{z})^2 \theta(\hat{z})^2 \right) - \frac{k^2}{12 \hat{z}^3} \left( - \frac{3}{2} \theta(\hat{z})^2 \right) \right) \ .
\end{equation}
Following refs.~\cite{Karch:2005ms,Karch:2006bv} and evidence we will
present later, we propose the following $\theta$--dependent
counterterms:
\begin{equation}
S_{CT} = - 2 \pi^2 R^4 \mu_7 \int d^4 x \left( \frac{1}{2} \sqrt{-\gamma_4} \theta(\hat{\epsilon})^2 - \frac{5}{12} \sqrt{-\gamma_4} \theta(\hat{\epsilon})^4 - \frac{2}{3} \sqrt{-\gamma_4} \chi(\hat{\epsilon})^2 \theta(\hat{\epsilon})^2 \ln ( \theta( \hat{\epsilon})) \right) \ ,
\end{equation}
where the first two terms are from
refs.~\cite{Karch:2005ms,Karch:2006bv} and the third term is motivated
by the logarithm term in ref.~\cite{Karch:2006bv}.  Furthermore,
$\gamma_4$ is the four dimensional metric at the AdS boundary such that:
\begin{equation}
\sqrt{-\gamma_4} = \frac{1}{16 \hat{\epsilon}^4} - \frac{k^2}{12 \hat{\epsilon}^2} \ .
\end{equation}
We emphasize that one has to keep the second divergent term (this is
also true to derive the results of ref.~\cite{Karch:2006bv}).

\subsection{Extracting the Condensate}
Following ref.~\cite{Karch:2005ms} to calculate the condensate, we
first Euclideanize our action
and the (regularized) action is simply the sum of the Euclideanized bare and counterterm actions:
\begin{equation}
I_R=  I_{D7} + I_{CT}  \ , \quad \mathcal{F} = \frac{I_R}{\beta} \ ,
\end{equation}
where $\beta$ is the period of the periodic time $\tau=-it$. Then we
can calculate the condensate (density) as:
\begin{equation}
\langle \mathcal{O} \rangle = - \frac{1}{V} \frac{ \delta \mathcal{F}}{\delta m_q} =  - \frac{1}{\beta V} \frac{\delta I_R}{\delta m_q} \ .
\end{equation}
Given possible different normalizations of the volume term, we write:
\begin{equation}
\beta V = \hat{\epsilon}^4 \int d^4 x \sqrt{\gamma_4} \ .
\end{equation}
Furthermore, the bare quark mass is associated with the asymptotic
separation $L$ of the D7--brane from the stack of D3--branes:
\begin{equation}
m_q = \frac{L}{2 \pi \alpha'}  \rightarrow \delta m_q = \frac{1}{2 \pi \alpha'} \delta L \ .
\end{equation}
The asymptotic separation is given by
\begin{equation}
 L = \lim_{\hat{z} \to 0} \frac{R}{\hat{z}} \sin\theta(\hat{z}) = R \ \theta_0 \ ,
\end{equation}
so we have:
\begin{equation}
\delta L = \frac{R}{\hat{\epsilon}}\delta \theta(\hat{\epsilon}) \ ,
\end{equation}
for a final result of:
\begin{equation}
\langle \mathcal{O} \rangle = - \frac{R}{\hat{\epsilon}^3 \int d^4 x \sqrt{\gamma_4}} \frac{\delta I_{R}}{\delta \theta(\hat{\epsilon})} = 2 \pi^2 R^5 \mu_7 \left(- 2 \theta_2 + \frac{1}{3} \theta_0^3 - \frac{4}{3} k^2 \theta_0  \ln(\theta_0) \right)  \equiv 2 \pi^2 R^5 \mu_7 C \ .
\end{equation}
The new counterterm we have introduced is not the only term one can write down in order to cancel the UV logarithm divergence in the action.  Furthermore, we can write terms that are finite in the UV.  However, we find our counterterm is the simplest counterterm we can write that cancels the UV logarithm divergence in the action, has a condensate with no dependence on the UV cutoff $\epsilon$, as well as give the correct condensate asymptotics as we show later.
%
\subsection{Embeddings and the Singular Locus} \label{sec:embeddings}
%
Consider  the volume  of  the  3--sphere in  terms  of the  coordinate
$r'$.  This  volume  always   vanishes  at  $\theta=\pi/2$.  Away  from
$\theta=\pi/2$,  we find  that it  is finite  at $r_0'$  (the singular
locus) for $\gamma  = 0$, whereas it vanishes at  $r_0'$ for $\gamma <
0$.  

\begin{figure}[h] 
   \centering
   \includegraphics[width=3.0in]{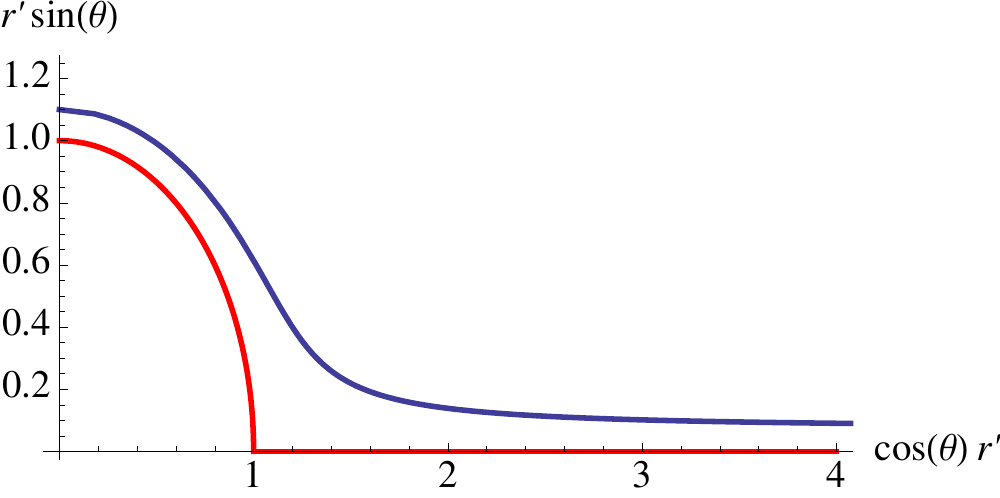} 
   \caption{\small Depiction of zero mass embedding with a finite mass embedding for $\gamma =0$.}  \label{fig:zeromass}
\end{figure}

Let us study each case separately.  For $\gamma = 0$, the result that
the 3--sphere volume is finite suggests that the embedding given by
the simple zero mass embedding solution $\theta(r') = 0$ which extends
from $r' = \infty$ to $r' = r_0' = R$ cannot simply end there.  If we
consider a D7--brane embedding such that $r(\theta)$ is the new
transverse direction, $r(\theta) = r_0'$ is a solution.  Connecting
this to the previous segment allows us to extend the zero mass
embedding up to $\theta = \pi/2$, where the 3--sphere volume shrinks
to zero, and the D7--brane can end.  We depict this embedding in
figure~\ref{fig:zeromass}, in units where $r'_0=R=1$. It is the solid
curve that runs along the horizontal axis ($\theta=0$) and then
follows the $r=1$ arc (the singular locus) to end at $\theta=\pi/2$,
the location of the enhan\c con. 
Embeddings with non--zero mass start at infinity at some non--zero
$\theta_0$, run roughly horizontally until deviating North, skirting
the singular locus and then ending on the $\theta=\pi/2$ axis. Such
trajectories show that the singular locus in the interior of the
geometry acts to repel the D7--branes somewhat\footnote{The enhan\c con geometry can be thought of as having an orientifold O7 with massive D7--branes \cite{HoyosBadajoz:2010td}.  It is possible the repulsion we see is due to the D7--brane charge.  We thank Carlos Hoyos for suggesting this.}. However, unlike other
cases where repulsion generates a condensate even for the zero mass
embedding giving spontaneous chiral symmetry breaking (e.g.,
background magnetic field \cite{Filev:2007gb}, see also the examples
in refs.~\cite{Babington:2003vm,Evans:2005ti}) the condensate at zero
mass remains vanishing. We depict an example in figure
\ref{fig:zeromass}. Very high masses give embeddings that are hardly
deviated by the singular locus, their behaviour being very similar to
that in ordinary AdS.

\begin{figure}[h] 
   \centering
   \includegraphics[width=3.0in]{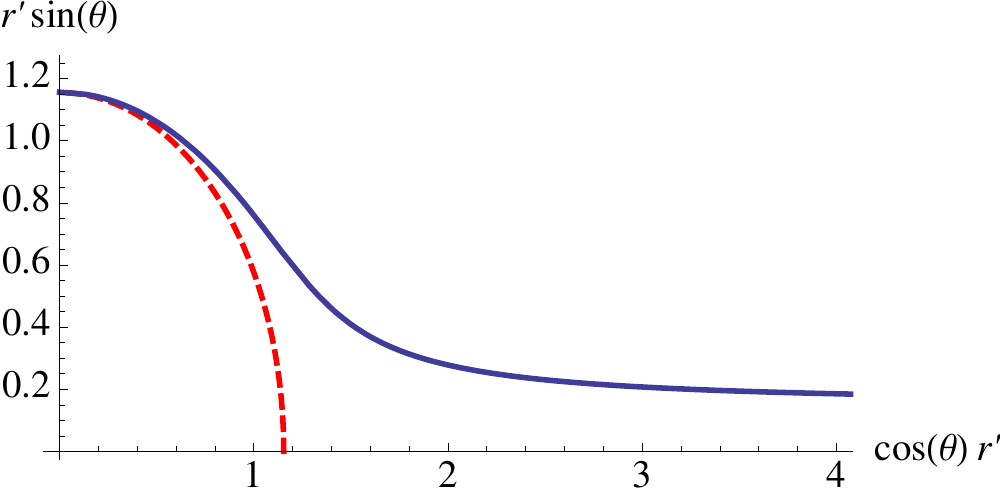} 
   \caption{\small Depiction of lowest mass embedding for $\gamma = -0.002$.  The red, dashed line is the singular locus.}  \label{fig:gammaembedding}
\end{figure}

\begin{figure}[h] 
   \centering
   \includegraphics[width=2.5in]{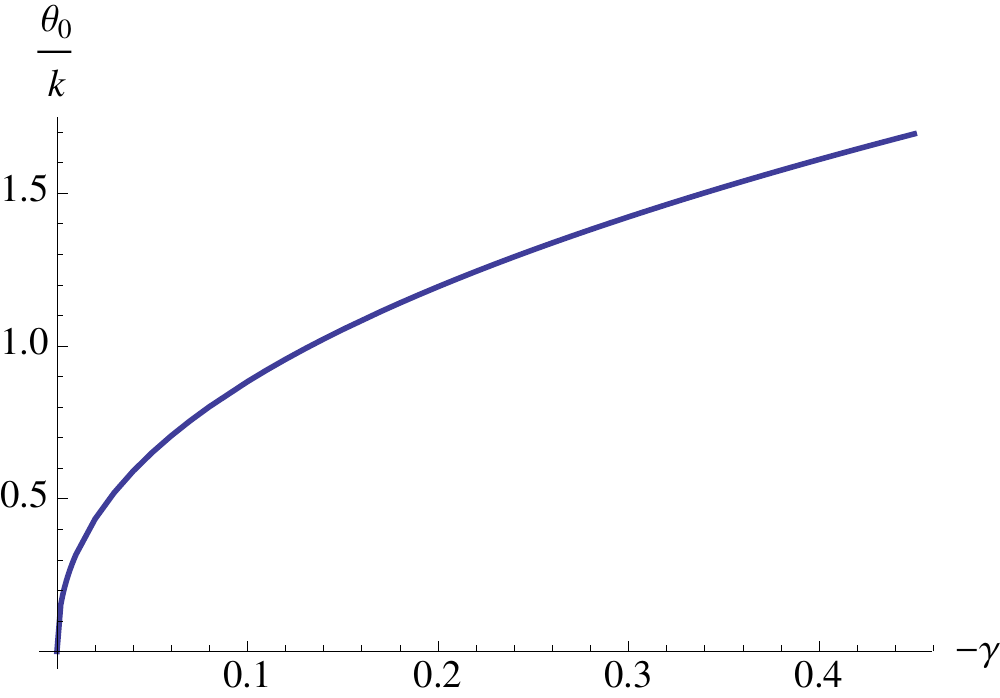} 
   \caption{\small Lowest mass as a function of $\gamma$.}  \label{fig:lowestmass}
\end{figure}

For $\gamma <0$, since the 3--sphere shrinks to zero size on the
singular locus, one might think that no extension is needed, with the
D7--brane ending at $r'_0$ and some value of $\theta$. This would be
analogous to the finite temperature case, with the singular locus
playing a role similar to the
horizon\cite{Babington:2003vm,Albash:2006ew,Mateos:2006nu}.  However,
a study of the induced metric on the D7--brane as they approach the
singular locus shows that embeddings that end on the singular locus
will have a conical singularity.  Therefore, for $\gamma < 0$, no
embeddings can end at $r = r_0'$, and they must run to some final
value of $r'$ at $\theta=\pi/2$.  An example of such a consistent
embedding is shown in figure~\ref{fig:gammaembedding}.  A direct
consequence of this result is that for $\gamma < 0$, no zero mass
embedding exists. There is in fact a minimum mass embedding, and as
$\gamma$ decreases, the minimum mass increases in value.  We show this
in figure~\ref{fig:lowestmass}, plotting the bare mass against
$-\gamma$. (Recall that $\theta_0=2\pi\alpha' m_q/R$.) For a given
non--zero $\gamma$, this ``gap'' in the spectrum of allowed masses is
set by the adjoint mass $k=mR$.

\subsection{Some Analytics} \label{sec:analytics}
%
Consider solving for the field $\theta(\hat{z})$ analytically as an
expansion in small $k$.  We proceed by defining a new field:
\begin{equation}
\Phi(\hat{z}) = \sin( \theta(\hat{z})) = f_0(\hat{z}) + k^2 f_1(\hat{z}) + \mathcal{O}(k^4) \ .
\end{equation}
Putting this into the equations of motion, and solving for each
function $f_i(\hat{z})$, we first find the well known solution for
pure AdS :
\begin{equation}
f_0(\hat{z}) = \hat{z} c_0 \ .
\end{equation}
For pure AdS, $c_0$ is related to the bare quark mass.  The function
$f_1(\hat{z})$ has solution given by:
\begin{equation}
f_1(\hat{z}) =  \frac{c_1 \hat{z} + c_2 \hat{z}^3+ \frac{1}{3} c_0 \hat{z}^3 -  c_0^3 \hat{z}^5 - \frac{2}{3} c_0 \hat{z}^3 \ln(\hat{z}) - \frac{4}{3} c_0^3 \hat{z}^5 \ln( k \hat{z}) }{ 1- c_0^2 \hat{z}^2 } \ ,
\end{equation}
where $(c_1, c_2)$ are constants.  These constants can be chosen based
on some dimensional analysis~\cite{Karch:2006bv} to be:
\begin{equation}
c_1 = - \frac{1}{3 c_0} \ , \quad c_2 = - \frac{2}{3} c_0 \ln(c_0) \ .
\end{equation}
Comparing to equation \reef{eqt:asymptotics}, we can extract the
asymptotic constants to be given by:
\begin{equation}
\theta_0 = c_0 - \frac{k^2}{3 c_0} \ , \quad  \theta_2 = \frac{c_0^3}{6} + k^2 \left(- \frac{c_0}{6} - \frac{2}{3} c_0 \ln(c_0) \right) \ .
\end{equation}
Substituting this into our condensate, we find that:
\begin{equation} \label{eqt:zeroc}
C  = 0 + \mathcal{O}(k^4) \ .
\end{equation}
Now, from dimensional analysis, the condensate
should have the following expansion in terms of $m/m_q$:
\begin{equation}
C = m_q^3 \left( o_0 + \left( \frac{m}{m_q} \right)^2 o_1 +  \left( \frac{m}{m_q} \right)^4 o_2 + \dots \right) \ ,
\end{equation}
for constants $o_i$.  Combining this with our result in equation
\reef{eqt:zeroc} (recall $k=mR$) tells us that the large mass limit of
the condensate is zero.

Another useful result to consider is how the mass and condensate
scale.  In particular, we note that the background only depends on the
combination $ k \hat{z}$, and therefore, we can absorb the constant
$k$ into a new coordinate $\tilde{z}$.  The asymptotic expansion for
$\theta(\tilde{z})$ is given by:
\begin{equation}
\theta(\tilde{z}) = \tilde{z} \left( \tilde{\theta}_0 + \tilde{z}^2 \left( \tilde{\theta}_2 - \frac{2}{3} \tilde{\theta}_0 \ln(\tilde{z}) \right) \right) \ .
\end{equation}
If we introduce the coordinate $\hat{z}$ into this expression, we find
that ({\it c.f.} equation~\reef{eqt:asymptotics}):
\begin{equation}
  \theta_0  = k \tilde{\theta}_0  \ , \quad \theta_2 = k^3 \left( \tilde{\theta}_2 - \frac{2}{3} \tilde{\theta}_0 \ln(k) \right) \ .
\end{equation}
Therefore, we learn that the bare quark mass should scale linearly with
$k$.  Substituting this result into our expression for the condensate, we
find that:
\begin{equation} \label{eqt:scaling}
C  = k^3 \tilde{C} = k^3 \left(- 2 \tilde{\theta}_2 + \frac{1}{3} \tilde{\theta}_0^3 - \frac{4}{3}  \tilde{\theta}_0  \ln(\tilde{\theta}_0) \right) \ .
\end{equation}
The condensate should scale as $k^3$. These results will be very
useful, as we shall see later.
%
\subsection{Numerical Regularization}
\begin{figure}[h] 
   \centering
   \includegraphics[width=2.5in]{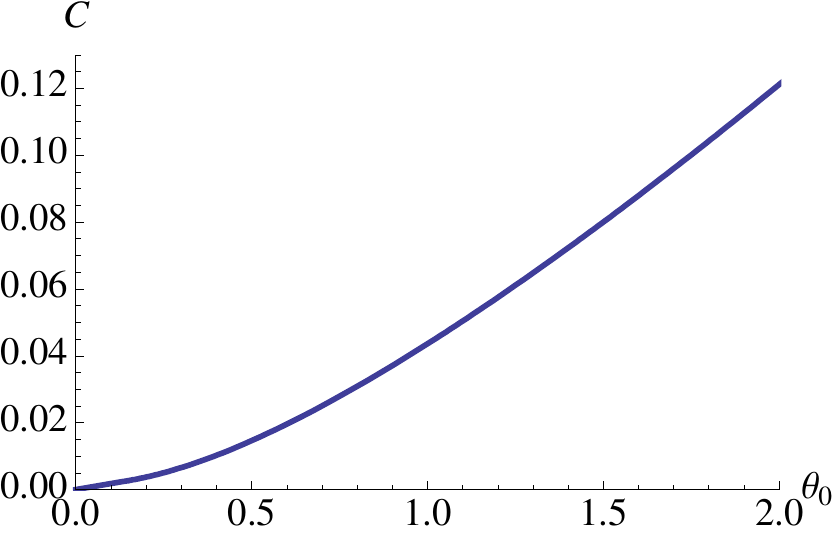} 
   \caption{\small Raw numerical data for the condensate as a function
     of bare quark mass for $k^2 = 10^{-2}$.  The deviation from zero
     for increasing mass is a numerical artifact, which we subtract in
     all later computations. Recall that $\theta_0=m_q\times
     (2\pi\alpha'/R)$.}
   \label{fig:c_vs_m_k=001}
\end{figure}

%
Before physics can be extracted from the numerical solution of the
non--linear equations of motion that result from the actions we write,
we must remove any systematic numerical effects that can occur in
regimes where accuracy is compromised by cancellation among very large
or very small numbers. Typically, these manifest themselves at high
bare quark mass in the condensate {\it vs.} mass curve, since in that
regime we are near the AdS boundary, where the conformal factor
naturally grows rapidly. We know from the previous section that the
condensate should fall to zero in this regime. The raw numerical
results plotted in figure~\ref{fig:c_vs_m_k=001} show that this is
masked by an offset. It is purely numerical and unphysical (it is very
similar to the deviation from zero seen for probe D7--branes in pure
AdS in global coordinates~\cite{Karch:2006bv}), and may be
consistently subtracted from all subsequent computations.  The
deviation from zero occurs for all values of $k$, and therefore, in
order to isolate the purely numerical piece to use as a regulator, we
choose to subtract off the curve for small $k$ (here we use $k =
10^{-1}$). After the appropriate rescaling given in
equation~\reef{eqt:scaling}, this curve can be used to extract the
physical condensate at higher $k$.  We stress again that this
subtraction is not physical in the sense that it is \emph {not}  akin to a
contribution of an unknown counterterm. We have treated those fully in
section~\ref{sec:counterterm}.

\begin{figure}[h]
\begin{center}
\subfigure[Not scaled]{\includegraphics[width=2.8in]{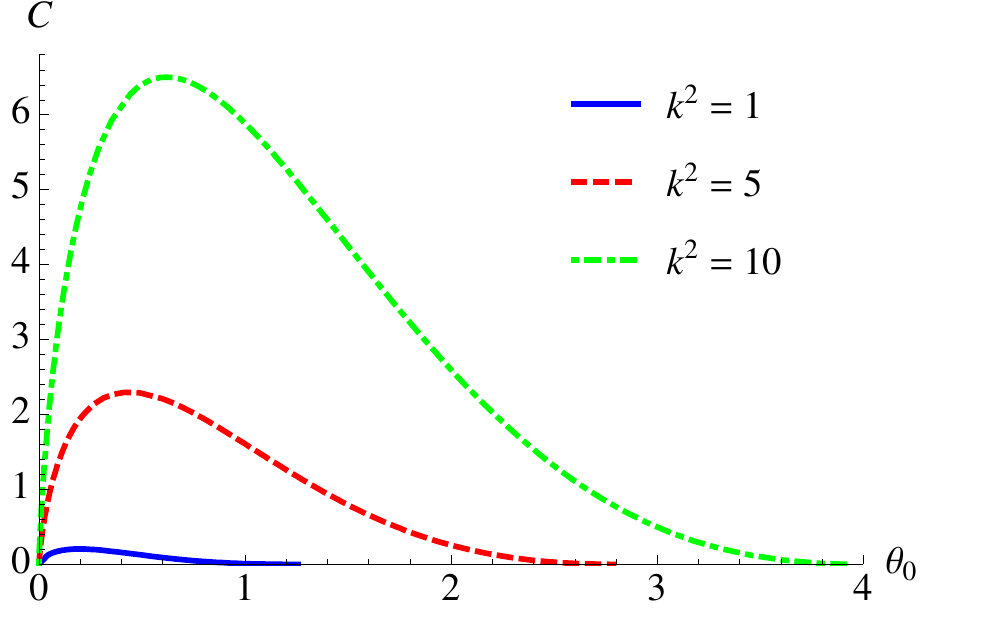}\label{fig:c_vs_m_unscaled}} \hspace{0.5cm}
\subfigure[Scaled]{\includegraphics[width=2.8in]{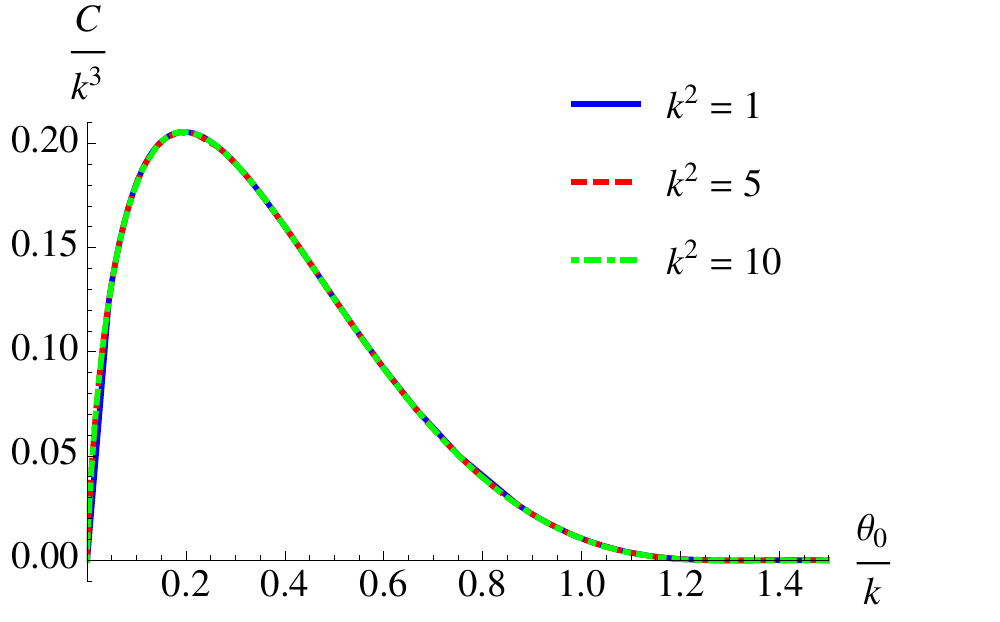}\label{fig:c_vs_m_scaled}}
   \caption{\small Condensate as a function of bare quark mass for $k^2 = 1, 5, 10$ for $\gamma = 0$. Recall that $\theta_0/k=m_q/m\times (2\pi\alpha'/R^2)$.}  \label{fig:c_vs_m}
   \end{center}
\end{figure}

\begin{figure}[h] 
   \centering
   \includegraphics[width=2.8in]{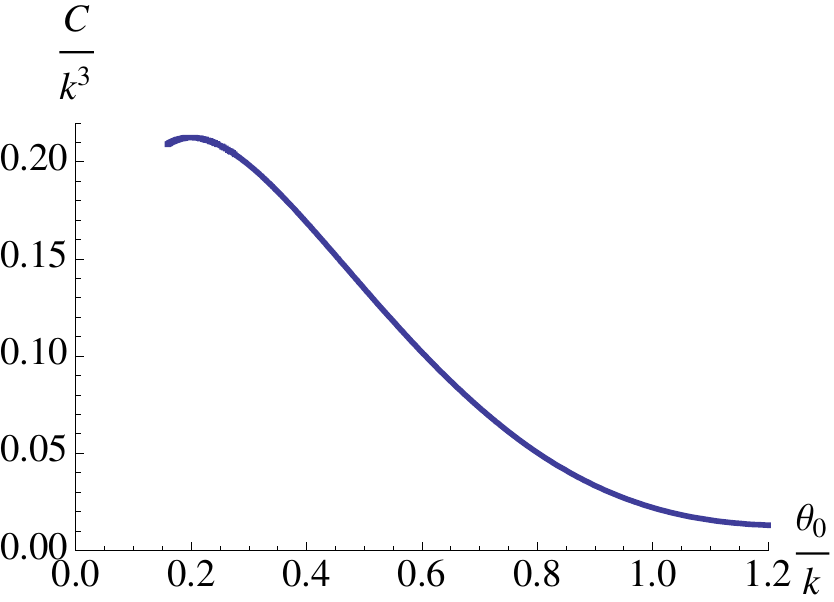} 
   \caption{\small Condensate as a function of bare quark mass for $\gamma = - 0.002$. Recall that $\theta_0/k=m_q/m\times (2\pi\alpha'/R^2)$.}
   \label{fig:c_vs_m_gamma}
\end{figure}

Using this procedure, we calculate the physical condensate in figure
\ref{fig:c_vs_m_unscaled}.  It is important to test that we have not
damaged the physics, and to that end we rescale each curve according
to equation \reef{eqt:scaling} and find they exactly match as shown in
figure \ref{fig:c_vs_m_scaled}.  We also present results for $\gamma <
0$, where we see that the absence of the states below a minimum mass
(see section~\ref{sec:embeddings}) simply removes a piece from the
condensate curve (see figure \ref{fig:c_vs_m_gamma}).
%
%
\section{Constituent Quark Mass} \label{sec:ConMass}
%
An important quantity to compute is the constituent quark mass. This
is given by the energy of a fundamental string stretching from the
D7--brane to the core of the geometry (made of D3--branes), starting
at the place the D7--brane ends.  This is to be contrasted with the
bare quark mass, which is the asymptotic separation of the D7--brane
from the D3--branes, extracted in the UV. The two types of mass are
readily visualized on figures~\ref{fig:zeromass}
and~\ref{fig:gammaembedding}. Far to the right the vertical distance
from embedding to the horizontal axis gives the bare quark mass. Far
to the left, on the $\theta=\pi/2$ axis, the vertical distance from
where the embedding ends to the singular locus gives the constituent
quark mass.

\begin{figure}[h] 
   \centering
   \includegraphics[width=2.8in]{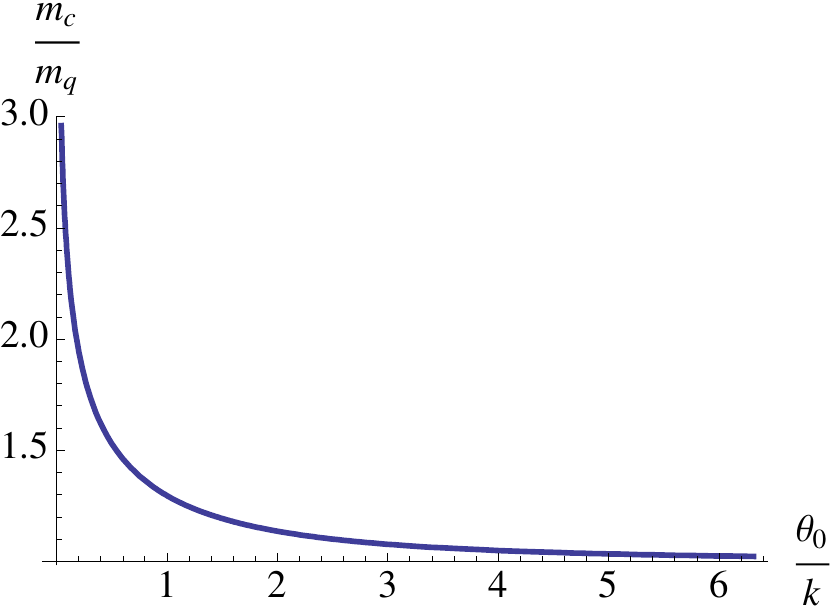} 
   \caption{\small Constituent to bare quark mass ratio {\it vs.} bare
     mass to adjoint mass ratio for the case $\gamma=0$. Recall that
     $\theta_0/k=m_q/m\times (2\pi\alpha'/R^2)$.}  \label{fig:ConMass}
\end{figure}

For the constituent quark mass, we calculate the energy of a string
that hangs from the point where the $S^3$ has shrunk to zero size (let
us denote this point as $z_m$) to $z = \infty$.  In order to do this,
we consider the Nambu--Goto action for the fundamental string (in
Einstein frame) given by:
\begin{equation}
  S = \frac{1}{2 \pi \alpha'} \int d^2 \sigma e^{\Phi/2} \sqrt{- h} = \frac{1}{2 \pi \alpha'}  \int d^2 \sigma  \mathcal{L} \ .
\end{equation}
Working in static gauge, we take as coordinates for the string
worldsheet $\sigma^0 = t$ and $\sigma^1 = z$.  In these coordinates,
the string does not couple to the NS--NS 2--form potential of the
background.  The physical energy (the constituent quark mass) is then
given by the action (per unit time) multiplied by a factor of two to
remove the factor of 1/2 from $g_{tt}$ near the AdS boundary):
\begin{equation}
\frac{2 \pi \alpha'}{R} m_c = 2 k \int_{z_m}^\infty dz \frac{ \rho^4 \coth(2 \chi) }{z} \ .
\end{equation}
It is convenient to use $\chi$ as an integration variable to write
(with the help of equations~\reef{eq:firstorder} and~\reef{eq:newz})
this equation as:
\begin{equation}
\frac{2 \pi \alpha'}{R} m_c = 4 k \int_{\chi_m}^\infty d\chi \  \mathrm{csch}(2 \chi) \coth(2 \chi) = 2 k \ \mathrm{csch}(2 \chi_m) \ .
\end{equation}
We present the results for this in figure~\ref{fig:ConMass}, for the
case $\gamma=0$.  The key result here is that the constituent mass is
always greater than the bare quark mass, and it is significantly
greater as the D7--brane approaches the singular locus.  This result
suggests that the embeddings experience a repulsive force from the
background, as we observed earlier.

\section{Meson Spectrum} \label{sec:meson}
\begin{figure}[h]
\begin{center}
\subfigure[Meson Spectrum]{\includegraphics[width=2.5in]{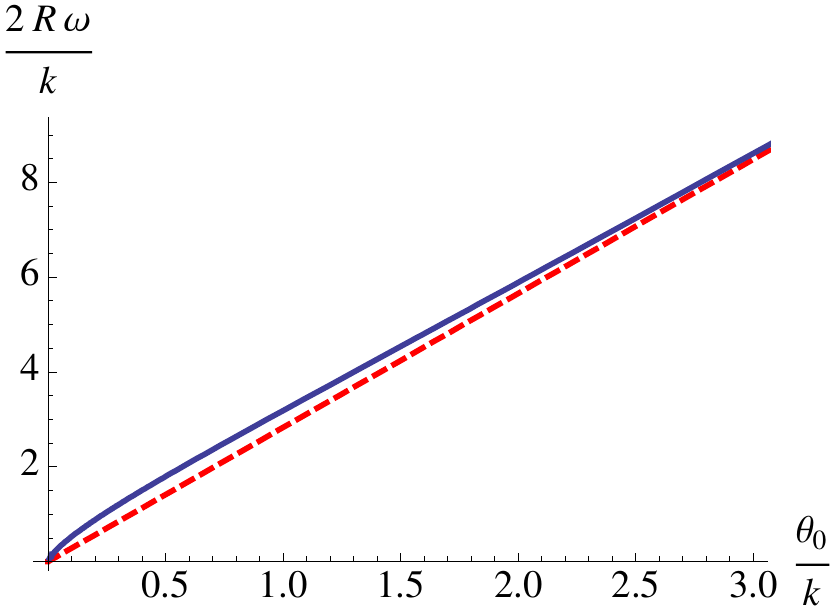}\label{fig:meson_nozoom}} \hspace{0.5cm}
\subfigure[Meson Spectrum Near Origin]{\includegraphics[width=2.5in]{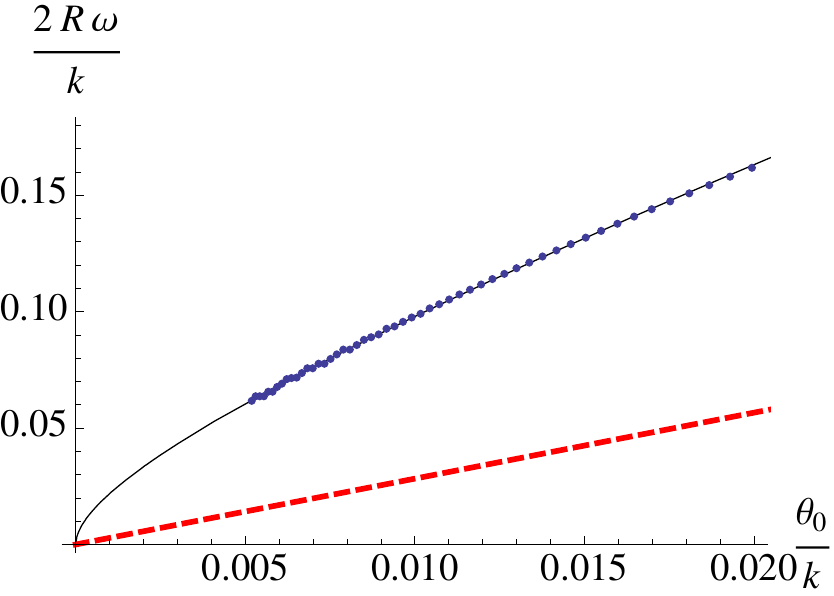}\label{fig:meson_zoom}}
\caption{\small Lowest level of the meson spectrum as a function of
  bare quark mass. (Recall that $\theta_0/k=m_q/m\times
  (2\pi\alpha'/R^2)$.)  The red--dashed line corresponds to the lowest
  level of the spectrum for $k=0$, \emph{i.e.} pure AdS$_5 \times S^5$
  studied in ref.~\cite{Kruczenski:2003be}.  In the figure to the
  right, the black line corresponds to the best fit
  curve.} \label{fig:meson}
 \end{center}
\end{figure}

%
We proceed to calculate part of the meson spectrum by calculating the
quadratic fluctuations associated with the field $\theta(z)$
\cite{Karch:2002xe}:
\begin{equation}
\theta(z) = \theta_0(z) + 2 \pi \alpha' \Phi(t, z)\ ,
\end{equation}
where $\theta_0(z)$ is the solution we found in the previous section.
Substituting this form into the Lagrangian and keeping only terms
quadratic in $\Phi$, we can calculate the equations of motion for the
field $\Phi(t,z)$.  Assuming a time dependence  of the form:
\begin{equation}
\Phi(t, z) = e^{- i \omega t} \phi(z) \ ,
\end{equation}
the asymptotic form  $g_{tt} \to \frac{R^2}{2
  z^2}$ means that $\omega$ is related to the mass of the meson via:
\begin{equation}
M = 2 \omega \ .
\end{equation}
Following a similar scaling analysis to that done in
section~\ref{sec:analytics}, we find that $\omega$ scales linearly
with~$k$, and therefore the relevant quantity for us will be the ratio
of the two.  We plot our results for the spectrum (extracted using
fairly standard techniques presented for example in
ref.\cite{Kruczenski:2003be}) in figure~\ref{fig:meson}.  As the
figure indicates, the spectrum approaches that of pure AdS as the bare
quark mass increases.  Furthermore, the mass of the meson is
consistently above that of pure AdS, which is consistent with our
previous result that the D7--brane feels a repulsive force.  Finally,
we find that the behavior of the curve near zero mass is well
approximated by a function of the following form:
\begin{equation}
\frac{2 R}{k} \omega = a \left(\frac{\theta_0}{k} \right)^{1/2} + b \frac{\theta_0}{k} \ ,
\end{equation}
where $(a,b)$ are constants.  This is reminiscent of the form of the
GMOR relation~\cite{Gell-Mann:1968rz}, although there is no spontaneous
chiral symmetry breaking here. It is worth exploring further the
origin of this behaviour, including how strongly it depends upon us
having restricted to $\phi = (2n+1)\pi/2$.

We consider separately the calculation of the meson spectrum for our
zero mass embedding.  We only need to consider the fluctuations of the
$\theta = 0$ embedding to extract this.  Following the same procedure
as before (but now in terms of the coordinate $r'$), we find near the
singular locus the solution for the fluctuation goes as:
\begin{equation}
\phi(r') \propto (r' - 1)^{-2} e^{\frac{\pm 3i  \hat{\omega}}{2(r'/R-1)}} \ ,
\end{equation}
where $\hat{\omega} = R \omega / k$.  This behavior is exactly the
ingoing/outgoing wave behavior we would expect for fluctuations near
the event horizon of an extremal black hole \cite{Kruczenski:2003be}.
Since we want our field to satisfy ingoing wave boundary conditions
(the plus sign), we can define a new field:
\begin{equation}
\phi(r') =  (r' - 1)^{-2} e^{\frac{ 3i  \hat{\omega}}{2(r'/R-1)}} y(r') \ .
\end{equation}
Near the AdS boundary, $y(r')$ behaves as:
\begin{equation}
y(r') \to c_1 r' + \frac{c_2}{r'} \ ,
\end{equation}
and we require that $c_1 = 0$ such that we only have a normalizable
mode.  We find that for a starting position of $r' = 1+ \epsilon$, the
value of $\hat{\omega}$ that satisfies the normalizibility condition
is given by $\hat{\omega} \approx - i \epsilon 7.7 \times 10^{-4} $
(starting with $\epsilon = 10^{-3})$.  Sending $\epsilon$ to zero gives
us $\hat{\omega} = 0$, which means we have a (stable) zero mass meson
for zero bare quark mass.
%
\section{Conclusions}\label{sec:conclusions}
%
  The results for our D7--brane probe analysis fit nicely with what had
  already been established for this background using D3--brane
  probes \cite{Buchel:2000cn,Evans:2000ct}. The enhan\c con locus, the special place the D3--branes
  become massless, is also the place where the D7--branes that are
  massless quark flavours end. All other D7--branes end above that
  locus, and in fact are somewhat bent away from it in their profile. 

  It this way we have confirmed the prediction of
  ref.~\cite{Evans:2005ti} that there is no chiral symmetry breaking
  in this background (although they seem to have thought that this
  form of the embedding would depend on one of the angles of the
  $S^3$, which we show is not necessary), since the zero mass
  embedding is not repelled by the enhan\c con (hence no chiral
  condensate). The higher mass embeddings do have a condensate
  associated with them of course, and we have explored the dependence
  on mass, finding it to be smooth, regardless of the value of the
  adjoint mass.  Much the same can be said about the accompanying
  spectrum of mesons we explored. 

  We have not addressed the meaning of our particular choices for the
  value of $\phi$, which allowed us to find the solutions. $\phi$ is
  related to the phase angle of the mass matrix in the dual theory
  \cite{Albash:2006bs}, and so it seems that this particular choice
  may have some physical meaning, but we have not yet explored that.  

  An important question to address is whether these choices are stable
  under fluctuations. Unfortunately, due to the complicated nature of
  the $C_{(6)}$ and $C_{(8)}$, we are not able to address this
  question by considering fluctuations of the field $\phi$ around our
  special values. However, it is important to note that the massless
  embedding is special here. The $\phi$--dependence simplifies even
  further at $\theta=0$, vanishing entirely. This suggests that the
  zero mass embedding persists for any value of $\phi$, as might have
  been expected, and that there are additional symmetries in that
  case. Possibly the embedding itself is supersymmetric, but in
  addition there is probably a continuous symmetry (translations in
  $\phi$) that appears for that sector. This may also be connected to
  the GMOR--like behaviour that we observed for the meson mass
  dependence as the bare quark mass approaches zero, although that
  meson is from the $\theta$ fluctuations, and we do not have a
  spontaneous symmetry breaking here.

  There appears to be some qualitative similarities between features
  of our solution and that of solutions in global AdS.  It seems that
  our parameter $k=mR$, the adjoint mass, plays a similar role to the
  inverse radius of the $S^3$ on which the dual theory lives.  A
  similar modification to the asymptotics in the field $\theta$
  occurs, so we find a similar counterterm is needed.  It would be
  nice to understand this further. The type of counterterms that can
  arise seem connected to the way in which the (non-thermal)
  parameter/deformation under discussion is introduced. Sometimes it
  is at the level of the background itself (such as in the global
  case, of the case here) while in the case of a finite magnetic field
  \cite{Filev:2007gb} or chemical potential \cite{Karch:2006bv} where
  there is no logarithmic behavior, the parameter is introduced at the
  level at the level of the D7--brane action itself. This is worth
  further exploration.
\section*{Acknowledgments}
This work was supported by the US Department of Energy and the USC College of Letters, Arts, and Sciences.
\providecommand{\href}[2]{#2}\begingroup\raggedright\endgroup


\end{document}